\documentclass[a4paper,10pt]{article}
\usepackage{RRA4}

\usepackage{hyperref}
\usepackage[latin1]{inputenc}
\usepackage{graphicx}
\usepackage{amsmath}
\usepackage{amsfonts}
\usepackage{amssymb}
\usepackage{makeidx}
\usepackage{url}
\usepackage{color}
\usepackage{latexsym}
\usepackage{subfig}
\usepackage{paralist}

\RRdate{Octobre 2009}
\RRauthor{
Marine Minier
  \and
Benjamin Pousse

}
\authorhead{Minier \& Pousse}
\RRtitle{Amélioration des attaques intégrales contre Rijndael}
\RRetitle{Improving Integral Cryptanalysis against Rijndael with Large Blocks}
\titlehead{}
\RRresume{Ce rapport présente de nouvelles propriétés intégrales pour les variantes de Rijndael pour des blocs de tailles supérieures à 128 bits. En utilisant des distingueurs particuliers et des extensions d'attaques connues, les propriétés déduites permettent d'attaquer 7 et 8 étages de Rijndael.
}
\RRabstract{This report presents new four-round integral properties against the Rijndael cipher with block sizes larger than 128 bits. Using higher-order multiset distinguishers and other well-known extensions of those properties, the deduced attacks reach up to 7 and 8 rounds of Rijndael variants with 160 up to 256-bit blocks. For example, a 7-rounds attack against Rijndael-224 has a time complexity equal to $2^{80}$.}
\RRmotcle{chiffrement par blocs, cryptanalyses, attaques intégrales, Rijndael-$b$}
\RRkeyword{block cipher, cryptanalysis, integral attacks, Rijndael-$b$}
\RRprojets{SWING}
\RRtheme{\THCom} 
 \URRhoneAlpes 
\begin{document}
\makeRR   

\section{Introduction}

Rijndael-$b$ is an SPN block cipher designed by Vincent Rijmen and Joan Daemen \cite{Rijndael02}. It has been chosen as the new advanced encryption standard by the NIST \cite{AES99} with a 128-bit block size and a variable key length, which can be set to
128, 192 or 256 bits. In its full version, the block lengths $b$ and the key lengths $Nk$ can range from 128 up to 256 bits in steps of 32 bits, as detailed in \cite{DR02} and in \cite{DBLP:conf/mycrypt/NakaharaFP05}. There are 25 instances of Rijndael. The number of rounds $Nr$ depends on the text size $b$ and on the key size $Nk$ and varies between 10 and 14 (see Table \ref{sr} for partial details). For all the versions, the current block at the input of the round $r$ is represented by a $4 \times t$ with $t=(b/32)$ matrix of bytes $A^{(r)}$:

$$
A^{(r)} =
\left(
\begin{array}{cccc}
a^{(r)}_{0,0} & a^{(r)}_{0,1} & \cdots & a^{(r)}_{0,t}\\
a^{(r)}_{1,0} & a^{(r)}_{1,1} & \cdots &  a^{(r)}_{1,t} \\
a^{(r)}_{2,0} & a^{(r)}_{2,1} & \cdots & a^{(r)}_{2,t} \\
a^{(r)}_{3,0} & a^{(r)}_{3,1} & \cdots & a^{(r)}_{3,t} \\
\end{array}
\right)
$$

The round function, repeated $Nr-1$ times, involves four elementary
mappings, all linear except the first one:
\begin{compactitem}
\item SubBytes: a bytewise transformation that applies on each byte of the current block an 8-bit to 8-bit non linear S-box $S$.
\item ShiftRows: a linear mapping that rotates on the left all the rows of the current matrix. the values of the shifts (given in Table \ref{sr}) depend on $b$.
\item MixColumns: a linear matrix multiplication; each column of the input matrix is multiplied by the matrix $M$ that provides the corresponding column of the output matrix. 
\item AddRoundKey: an x-or between the current block and the subkey of the round $r$ $K_r$. 
\end{compactitem}

Those $Nr-1$ rounds are surrounded at the top by an initial key
addition with the subkey $K_0$ and at the bottom by a final
transformation composed by a call to the round function where the
MixColumns operation is omitted. The key schedule derives $Nr + 1$ $b$-bits round keys $K_0$ to $K_{Nr}$ from the master key $K$ of variable length.

\vspace{-3mm}
\begin{table}[ht]
\begin{center}
\begin{tabular}{c|ccccc}
& AES & Rijndael-160 & Rijndael-192 & Rijndael-224 & Rijndael-256 \\
\hline
ShiftRows & (1,2,3) & (1,2,3) & (1,2,3) & (1,2,4) & (1,3,4) \\
\hline
$Nb$ rounds ($Nk$=128) & 10 & 11 & 12 & 13 & 14 \\ 
\hline
$Nb$ rounds ($Nk$=192) & 12 & 12 & 12 & 13 & 14 \\
\hline
$Nb$ rounds ($Nk$=256) & 14 & 14 & 14 & 14 & 14 \\
\end{tabular}
\caption{Parameters of the Rijndael block cipher where the triplet $(i,j,k)$ for the ShiftRows operation designated the required number of byte shifts for the second row, the third one and the fourth one.} \label{sr}
\end{center}
\end{table}
\vspace{-4mm}

Many cryptanalyses have been proposed against Rijndael-$b$, the first one against all the versions of Rijndael-$b$ is due to the algorithm designers themselves and is based upon integral properties (\cite{DKR97}, \cite{DR98}, \cite{DBLP:conf/fse/KnudsenW02}) that allows to efficiently distinguish 3 Rijndael inner rounds from a random permutation. This attack has been improved by Ferguson et al. in \cite{DBLP:conf/fse/FergusonKLSSWW00} allowing to cryptanalyse an 8 rounds version of Rijndael-$b$ with a complexity equal to $2^{204}$ trial encryptions and $2^{128}- 2^{119}$ plaintexts. 

Following the dedicated work of \cite{DBLP:conf/africacrypt/GaliceM08}, this paper presents new four-round integral properties of Rijndael-$b$ and the resulting 7 and 8 rounds attacks which are substantially faster than
exhaustive key search. Note also that those attacks greatly improve the previous results on Rijndael-$b$, essentially the ones given in \cite{DBLP:conf/mycrypt/NakaharaFP05} and in \cite{DBLP:conf/isw/NakaharaP07}.

This paper is organized as follows: Section \ref{sat} recalls the integral properties known against Rijndael-$b$ and investigates the new four and five rounds properties. Section \ref{attacks} presents the deduced 7 and 8 rounds attacks. Section \ref{conclusion} concludes this paper.
\vspace{-1mm}
\section{The integral properties} \label{sat}
We describe in this section the four inner rounds original integral property against the AES described in \cite{DBLP:conf/fse/FergusonKLSSWW00}, the five rounds integral property of Rijndael-256 described in \cite{DBLP:conf/africacrypt/GaliceM08} and the new integral properties against all the Rijndael-$b$ versions where $b$ is larger than 128 bits.

\subsection{Introduction and Notations}
In \cite{DBLP:conf/fse/KnudsenW02}, L. Knudsen and D. Wagner analyse integral cryptanalysis as a dual to differential attacks particularly applicable to block ciphers with bijective components. A first-order integral cryptanalysis considers a particular collection of $m$ words in the plaintexts and ciphertexts that differ on a particular component. The aim of this attack is thus to predict the values in the sums (i.e. the integral) of the chosen words after a certain number of rounds of encryption. The same authors also generalize this approach to higher-order integrals: the original set to consider becomes a set of $m^d$ vectors which differ in $d$ components and where the sum of this set is predictable after a certain number of rounds. The sum of this set is called a $d$th-order integral.

\subsubsection{Notations}
We first introduce and extend the consistent notations proposed in \cite{DBLP:conf/fse/KnudsenW02} for expressing word-oriented integral attacks. For a first order integral, we have:

\begin{compactitem}
\item The symbol `$\mathcal{C}$' (for ``Constant'') in the $i$th entry, means that the values of all the $i$th words in the collection of texts are equal. 
\item The symbol `$\mathcal{A}$' (for ``All'') means that all words in the collection of texts are different.
\item The symbol `$\mathcal{S}$' (for ``Sum'') means that the sum of all $i$th words can be predicted.
\item The symbol `$?$' means that the sum of words can not be predicted.
\end{compactitem}
For a $d$th order integral cryptanalysis:
\begin{compactitem}
\item The symbol `$\mathcal{A}^d$' corresponds with the components that participate in a $d$th-order integral, i.e. if a word can take $m$ different values then $\mathcal{A}^d$ means that in the integral, the particular word takes all values exactly $m^{d-1}$ times. 
\item The term `$A^d_i$' means that in the integral the string concatenation of all words with subscript $i$ take the $m^d$ values exactly once.
\item The symbol `$(\mathcal{A}^d_i)^k$' means that in the integral the string concatenation of all words with subscript $i$ take the $m^d$ values exactly $k$ times. 
\item The symbol `$Eq_i$' found for two different words means that the sums of all values taken on those particular words are equal. 
\end{compactitem}

\subsubsection{Integral properties of the AES}
\begin{figure}[ht]
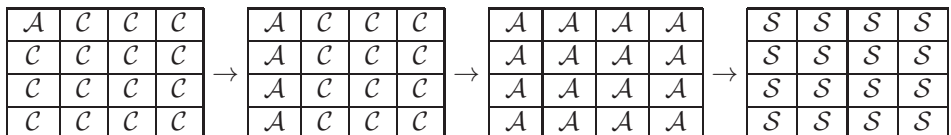

\begin{center}
\begin{tabular}{|c|c|c|c|}
\hline
$\mathcal{A}$ & $\mathcal{C}$ & $\mathcal{C}$ & $\mathcal{C}$ \\
\hline
$\mathcal{C}$ & $\mathcal{C}$ & $\mathcal{C}$ & $\mathcal{C}$ \\
\hline
$\mathcal{C}$ & $\mathcal{C}$ & $\mathcal{C}$ & $\mathcal{C}$ \\
\hline
$\mathcal{C}$ & $\mathcal{C}$ & $\mathcal{C}$ & $\mathcal{C}$ \\
\hline
\end{tabular}
$\rightarrow$
\begin{tabular}{|c|c|c|c|}
\hline
$\mathcal{A}$ & $\mathcal{C}$ & $\mathcal{C}$ & $\mathcal{C}$ \\
\hline
$\mathcal{A}$ & $\mathcal{C}$ & $\mathcal{C}$ & $\mathcal{C}$ \\
\hline
$\mathcal{A}$ & $\mathcal{C}$ & $\mathcal{C}$ & $\mathcal{C}$ \\
\hline
$\mathcal{A}$ & $\mathcal{C}$ & $\mathcal{C}$ & $\mathcal{C}$ \\
\hline
\end{tabular}
$\rightarrow$
\begin{tabular}{|c|c|c|c|}
\hline
$\mathcal{A}$ & $\mathcal{A}$ & $\mathcal{A}$ & $\mathcal{A}$ \\
\hline
$\mathcal{A}$ & $\mathcal{A}$ & $\mathcal{A}$ & $\mathcal{A}$ \\
\hline
$\mathcal{A}$ & $\mathcal{A}$ & $\mathcal{A}$ & $\mathcal{A}$ \\
\hline
$\mathcal{A}$ & $\mathcal{A}$ & $\mathcal{A}$ & $\mathcal{A}$ \\
\hline
\end{tabular}
$\rightarrow$
\begin{tabular}{|c|c|c|c|}
\hline
$\mathcal{S}$ & $\mathcal{S}$ & $\mathcal{S}$ & $\mathcal{S}$ \\
\hline
$\mathcal{S}$ & $\mathcal{S}$ & $\mathcal{S}$ & $\mathcal{S}$ \\
\hline
$\mathcal{S}$ & $\mathcal{S}$ & $\mathcal{S}$ & $\mathcal{S}$ \\
\hline
$\mathcal{S}$ & $\mathcal{S}$ & $\mathcal{S}$ & $\mathcal{S}$ \\
\hline
\end{tabular}
\caption{The 3-round first-order integral for Rijndael-128, where $\mathcal{S} = 0$} \label{int}
\end{center}
\end{figure}

In order to well understand the principles of an integral property, we give the example of the AES (Rijndael-128). Consider a collection of 256 texts, which have different values in one byte and equal values in all other bytes. Then it follows that after two rounds of encryption
the texts take all 256 values in each of the sixteen bytes, and that after three
rounds of encryption the sum of the 256 bytes in each position is zero as shown in \cite{DR98}. Also, note that there are 16 such integrals since the position of the non-constant byte
in the plaintexts can be in any of the sixteen bytes. The integral is illustrated in
Figure \ref{int} (where an arrow represents a complete round). This integral can be used to attack four rounds of Rijndael-128 with small complexity (note that the final round is special and does
not include MixColumns) counting over one key byte at a time. Simply guess a
key byte and compute byte-wise backwards to check if the sum of all 256 values
is zero.

\begin{figure}[ht]
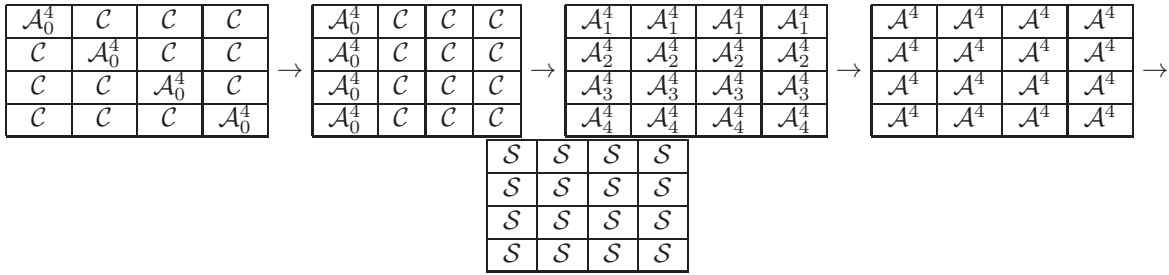

\begin{center}
\begin{tabular}{|c|c|c|c|}
\hline
$\mathcal{A}^4_0$ & $\mathcal{C}$ & $\mathcal{C}$ & $\mathcal{C}$ \\
\hline
$\mathcal{C}$ & $\mathcal{A}^4_0$ & $\mathcal{C}$ & $\mathcal{C}$ \\
\hline
$\mathcal{C}$ & $\mathcal{C}$ & $\mathcal{A}^4_0$ & $\mathcal{C}$ \\
\hline
$\mathcal{C}$ & $\mathcal{C}$ & $\mathcal{C}$ & $\mathcal{A}^4_0$ \\
\hline
\end{tabular}
$\rightarrow$
\begin{tabular}{|c|c|c|c|}
\hline
$\mathcal{A}^4_0$ & $\mathcal{C}$ & $\mathcal{C}$ & $\mathcal{C}$ \\
\hline
$\mathcal{A}^4_0$ & $\mathcal{C}$ & $\mathcal{C}$ & $\mathcal{C}$ \\
\hline
$\mathcal{A}^4_0$ & $\mathcal{C}$ & $\mathcal{C}$ & $\mathcal{C}$ \\
\hline
$\mathcal{A}^4_0$ & $\mathcal{C}$ & $\mathcal{C}$ & $\mathcal{C}$ \\
\hline
\end{tabular}
$\rightarrow$
\begin{tabular}{|c|c|c|c|}
\hline
$\mathcal{A}^4_1$ & $\mathcal{A}^4_1$ & $\mathcal{A}^4_1$ & $\mathcal{A}^4_1$ \\
\hline
$\mathcal{A}^4_2$ & $\mathcal{A}^4_2$ & $\mathcal{A}^4_2$ & $\mathcal{A}^4_2$ \\
\hline
$\mathcal{A}^4_3$ & $\mathcal{A}^4_3$ & $\mathcal{A}^4_3$ & $\mathcal{A}^4_3$ \\
\hline
$\mathcal{A}^4_4$ & $\mathcal{A}^4_4$ & $\mathcal{A}^4_4$ & $\mathcal{A}^4_4$ \\
\hline
\end{tabular}
$\rightarrow$
\begin{tabular}{|c|c|c|c|}
\hline
$\mathcal{A}^4$ & $\mathcal{A}^4$ & $\mathcal{A}^4$ & $\mathcal{A}^4$ \\
\hline
$\mathcal{A}^4$ & $\mathcal{A}^4$ & $\mathcal{A}^4$ & $\mathcal{A}^4$ \\
\hline
$\mathcal{A}^4$ & $\mathcal{A}^4$ & $\mathcal{A}^4$ & $\mathcal{A}^4$ \\
\hline
$\mathcal{A}^4$ & $\mathcal{A}^4$ & $\mathcal{A}^4$ & $\mathcal{A}^4$ \\
\hline
\end{tabular}
$\rightarrow$
\begin{tabular}{|c|c|c|c|}
\hline
$\mathcal{S}$ & $\mathcal{S}$ & $\mathcal{S}$ & $\mathcal{S}$ \\
\hline
$\mathcal{S}$ & $\mathcal{S}$ & $\mathcal{S}$ & $\mathcal{S}$ \\
\hline
$\mathcal{S}$ & $\mathcal{S}$ & $\mathcal{S}$ & $\mathcal{S}$ \\
\hline
$\mathcal{S}$ & $\mathcal{S}$ & $\mathcal{S}$ & $\mathcal{S}$ \\
\hline
\end{tabular}
\caption{A 4-round fourth-order integral for Rijndael-128 with $2^{32}$ texts.} \label{int4}
\end{center}
\end{figure}

This 3-round property could be extended by one round at the beginning using $2^{32}$ plaintexts, i.e. to a 4th-order integral property as described in \cite{DBLP:conf/fse/FergusonKLSSWW00}. The main observation is that the
$2^{32}$ plaintexts could be seen as $2^{24}$ copies of the above first-order integrals (starting in the second round). Since the text in each integral sums to zero in any byte after the fourth round, so does the sum of all $2^{32}$ plaintexts. The running time complexity of this attack greatly improves the previous one especially concerning the key-bytes search. Figure \ref{int4} depicts this four-round fourth-order integral against Rijndael-128.

\vspace{-3mm}
\vspace{-3mm}

\vspace{-3mm}
\vspace{-3mm}

\subsubsection{An integral property for Rijndael-256}
\begin{figure}[ht]
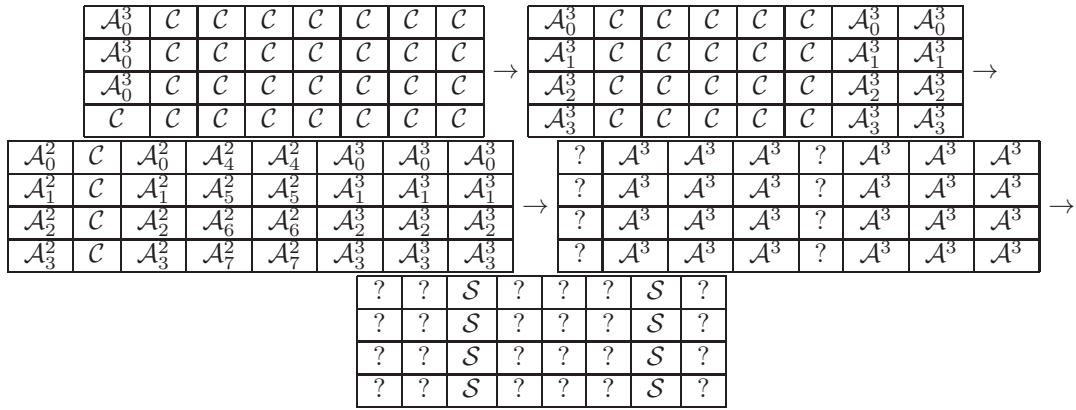

\begin{center}
\begin{tabular}{|c|c|c|c|c|c|c|c|}
\hline
$\mathcal{A}^3_0$ & $\mathcal{C}$ & $\mathcal{C}$ & $\mathcal{C}$ & $\mathcal{C}$ & $\mathcal{C}$ & $\mathcal{C}$ & $\mathcal{C}$ \\
\hline
 $\mathcal{A}^3_0$ & $\mathcal{C}$ & $\mathcal{C}$ & $\mathcal{C}$ & $\mathcal{C}$ & $\mathcal{C}$ & $\mathcal{C}$ & $\mathcal{C}$ \\
\hline
 $\mathcal{A}^3_0$ & $\mathcal{C}$ & $\mathcal{C}$ & $\mathcal{C}$ & $\mathcal{C}$ & $\mathcal{C}$ & $\mathcal{C}$ & $\mathcal{C}$ \\
\hline
$\mathcal{C}$ & $\mathcal{C}$ & $\mathcal{C}$ & $\mathcal{C}$  & $\mathcal{C}$ & $\mathcal{C}$ & $\mathcal{C}$ & $\mathcal{C}$ \\
\hline
\end{tabular}
$\rightarrow$
\begin{tabular}{|c|c|c|c|c|c|c|c|}
\hline
$\mathcal{A}^3_0$ & $\mathcal{C}$ & $\mathcal{C}$ & $\mathcal{C}$ & $\mathcal{C}$ & $\mathcal{C}$ & $\mathcal{A}^3_0$ & $\mathcal{A}^3_0$ \\
\hline
 $\mathcal{A}^3_1$ & $\mathcal{C}$ & $\mathcal{C}$ & $\mathcal{C}$ & $\mathcal{C}$ & $\mathcal{C}$ & $\mathcal{A}^3_1$ & $\mathcal{A}^3_1$\\
\hline
$\mathcal{A}^3_2$ & $\mathcal{C}$ & $\mathcal{C}$ & $\mathcal{C}$ & $\mathcal{C}$ & $\mathcal{C}$ & $\mathcal{A}^3_2$ & $\mathcal{A}^3_2$ \\
\hline
$\mathcal{A}^3_3$ & $\mathcal{C}$ & $\mathcal{C}$  & $\mathcal{C}$ & $\mathcal{C}$ & $\mathcal{C}$ & $\mathcal{A}^3_3$ & $\mathcal{A}^3_3$ \\
\hline
\end{tabular}
$\rightarrow$
\begin{tabular}{|c|c|c|c|c|c|c|c|}
\hline
$\mathcal{A}^2_0$ & $\mathcal{C}$ & $\mathcal{A}^2_0$ & $\mathcal{A}^2_4$ & $\mathcal{A}^2_4$ & $\mathcal{A}^3_0$ & $\mathcal{A}^3_0$ & $\mathcal{A}^3_0$ \\
\hline
 $\mathcal{A}^2_1$ & $\mathcal{C}$ & $\mathcal{A}^2_1$ & $\mathcal{A}^2_5$ & $\mathcal{A}^2_5$ & $\mathcal{A}^3_1$ & $\mathcal{A}^3_1$ & $\mathcal{A}^3_1$\\
\hline
$\mathcal{A}^2_2$ & $\mathcal{C}$ & $\mathcal{A}^2_2$ & $\mathcal{A}^2_6$ & $\mathcal{A}^2_6$ & $\mathcal{A}^3_2$ & $\mathcal{A}^3_2$ & $\mathcal{A}^3_2$ \\
\hline
$\mathcal{A}^2_3$ & $\mathcal{C}$ & $\mathcal{A}^2_3$  & $\mathcal{A}^2_7$ & $\mathcal{A}^2_7$ & $\mathcal{A}^3_3$ & $\mathcal{A}^3_3$ & $\mathcal{A}^3_3$ \\
\hline
\end{tabular}
$\rightarrow$
\begin{tabular}{|c|c|c|c|c|c|c|c|}
\hline
$?$ & $\mathcal{A}^3$ & $\mathcal{A}^3$ & $\mathcal{A}^3$ & $?$ & $\mathcal{A}^3$ & $\mathcal{A}^3$ & $\mathcal{A}^3$ \\
\hline
 $?$ & $\mathcal{A}^3$ & $\mathcal{A}^3$ & $\mathcal{A}^3$ & $?$ & $\mathcal{A}^3$ & $\mathcal{A}^3$ & $\mathcal{A}^3$ \\
\hline
 $?$ & $\mathcal{A}^3$ & $\mathcal{A}^3$ & $\mathcal{A}^3$ & $?$ & $\mathcal{A}^3$ & $\mathcal{A}^3$ & $\mathcal{A}^3$ \\
\hline
$?$ & $\mathcal{A}^3$ & $\mathcal{A}^3$ & $\mathcal{A}^3$  & $?$ & $\mathcal{A}^3$ & $\mathcal{A}^3$ & $\mathcal{A}^3$ \\
\hline
\end{tabular}
$\rightarrow$
\begin{tabular}{|c|c|c|c|c|c|c|c|}
\hline
$?$ & $?$ & $\mathcal{S}$ & $?$ & $?$ & $?$ & $\mathcal{S}$ & $?$ \\
\hline
 $?$ & $?$ & $\mathcal{S}$ & $?$ & $?$ & $?$ & $\mathcal{S}$ & $?$ \\
\hline
 $?$ & $?$ & $\mathcal{S}$ & $?$ & $?$ & $?$ & $\mathcal{S}$ & $?$ \\
\hline
$?$ & $?$ & $\mathcal{S}$ & $?$  & $?$ & $?$ & $\mathcal{S}$ & $?$ \\
\hline
\end{tabular}
\end{center}
\caption{4-round 3th-order integral property of Rijndael-256} \label{256_1}
\end{figure}

In \cite{DBLP:conf/africacrypt/GaliceM08}, the authors show a new 3th-order integral property against 4-round of Rijndael-256 which essentially relies on the slow diffusion of Rijndael-256. Using the previous notations, we could easily describe this property as shown on Figure \ref{256_1}.

As shown in \cite{DBLP:conf/africacrypt/GaliceM08}, this particular property could be extended by one round at the beginning using a 4th-order integral (considering that it represents $2^{8}$ copies of the 3th-order four round integral) to build a 5-round distinguisher that uses $2^{32}$ plaintexts testing if the sum taken over all initial values of a particular byte belonging to the third or to the seventh column is equal to zero. This leads to the first 9-round attack against Rijndael-256.

\subsection{The new integral properties of Rijndael-$b$} 
In this section, we present new integral properties against Rijndael-$b$. Those properties have been found using always the same methodology: consider after one round a full column of active bytes, say $(y_0,y_1,y_2,y_3)$, then after two rounds, express each byte of the corresponding ciphertext according to this column. Thus, one could see the dependencies between two rounds bytes and can directly deduce the bytes that must take all possible values to obtain balanced bytes at the end of the third round and thus predictable sums at the end of the fourth round.

\subsubsection{Rijndael-256}
\begin{figure}[ht]
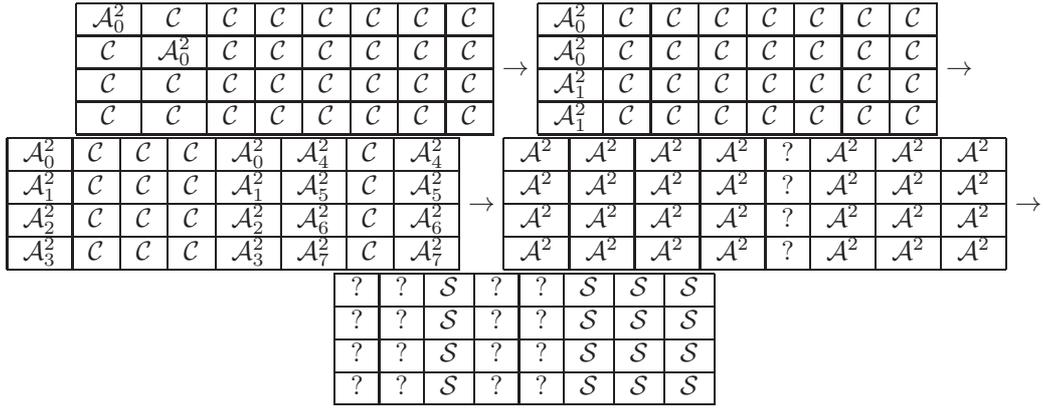

\begin{center}
\begin{tabular}{|c|c|c|c|c|c|c|c|}
\hline
$\mathcal{A}^2_0$ & $\mathcal{C}$ & $\mathcal{C}$ & $\mathcal{C}$ & $\mathcal{C}$ & $\mathcal{C}$ & $\mathcal{C}$ & $\mathcal{C}$ \\
\hline
 $\mathcal{C}$ & $\mathcal{A}^2_0$ & $\mathcal{C}$ & $\mathcal{C}$ & $\mathcal{C}$ & $\mathcal{C}$ & $\mathcal{C}$ & $\mathcal{C}$ \\
\hline
 $\mathcal{C}$ & $\mathcal{C}$ & $\mathcal{C}$ & $\mathcal{C}$ & $\mathcal{C}$ & $\mathcal{C}$ & $\mathcal{C}$ & $\mathcal{C}$ \\
\hline
$\mathcal{C}$ & $\mathcal{C}$ & $\mathcal{C}$ & $\mathcal{C}$  & $\mathcal{C}$ & $\mathcal{C}$ & $\mathcal{C}$ & $\mathcal{C}$ \\
\hline
\end{tabular}
$\rightarrow$
\begin{tabular}{|c|c|c|c|c|c|c|c|}
\hline
$\mathcal{A}^2_0$ & $\mathcal{C}$ & $\mathcal{C}$ & $\mathcal{C}$ & $\mathcal{C}$ & $\mathcal{C}$ & $\mathcal{C}$ & $\mathcal{C}$ \\
\hline
 $\mathcal{A}^2_0$ & $\mathcal{C}$ & $\mathcal{C}$ & $\mathcal{C}$ & $\mathcal{C}$ & $\mathcal{C}$ & $\mathcal{C}$ & $\mathcal{C}$\\
\hline
$\mathcal{A}^2_1$ & $\mathcal{C}$ & $\mathcal{C}$ & $\mathcal{C}$ & $\mathcal{C}$ & $\mathcal{C}$ & $\mathcal{C}$ & $\mathcal{C}$ \\
\hline
$\mathcal{A}^2_1$ & $\mathcal{C}$ & $\mathcal{C}$  & $\mathcal{C}$ & $\mathcal{C}$ & $\mathcal{C}$ & $\mathcal{C}$ & $\mathcal{C}$ \\
\hline
\end{tabular}
$\rightarrow$
\begin{tabular}{|c|c|c|c|c|c|c|c|}
\hline
$\mathcal{A}^2_0$ & $\mathcal{C}$ & $\mathcal{C}$ & $\mathcal{C}$ & $\mathcal{A}^2_0$ & $\mathcal{A}^2_4$ & $\mathcal{C}$ & $\mathcal{A}^2_4$ \\
\hline
 $\mathcal{A}^2_1$ & $\mathcal{C}$ & $\mathcal{C}$ & $\mathcal{C}$ & $\mathcal{A}^2_1$ & $\mathcal{A}^2_5$ & $\mathcal{C}$ & $\mathcal{A}^2_5$\\
\hline
$\mathcal{A}^2_2$ & $\mathcal{C}$ & $\mathcal{C}$ & $\mathcal{C}$ & $\mathcal{A}^2_2$ & $\mathcal{A}^2_6$ & $\mathcal{C}$ & $\mathcal{A}^2_6$ \\
\hline
$\mathcal{A}^2_3$ & $\mathcal{C}$ & $\mathcal{C}$  & $\mathcal{C}$ & $\mathcal{A}^2_3$ & $\mathcal{A}^2_7$ & $\mathcal{C}$ & $\mathcal{A}^2_7$ \\
\hline
\end{tabular}
$\rightarrow$
\begin{tabular}{|c|c|c|c|c|c|c|c|}
\hline
$\mathcal{A}^2$ & $\mathcal{A}^2$ & $\mathcal{A}^2$ & $\mathcal{A}^2$ & $?$ & $\mathcal{A}^2$ & $\mathcal{A}^2$ & $\mathcal{A}^2$ \\
\hline
 $\mathcal{A}^2$ & $\mathcal{A}^2$ & $\mathcal{A}^2$ & $\mathcal{A}^2$ & $?$ & $\mathcal{A}^2$ & $\mathcal{A}^2$ & $\mathcal{A}^2$ \\
\hline
 $\mathcal{A}^2$ & $\mathcal{A}^2$ & $\mathcal{A}^2$ & $\mathcal{A}^2$ & $?$ & $\mathcal{A}^2$ & $\mathcal{A}^2$ & $\mathcal{A}^2$ \\
\hline
$\mathcal{A}^2$ & $\mathcal{A}^2$ & $\mathcal{A}^2$ & $\mathcal{A}^2$  & $?$ & $\mathcal{A}^2$ & $\mathcal{A}^2$ & $\mathcal{A}^2$ \\
\hline
\end{tabular}
$\rightarrow$
\begin{tabular}{|c|c|c|c|c|c|c|c|}
\hline
$?$ & $?$ & $\mathcal{S}$ & $?$ & $?$ & $\mathcal{S}$ & $\mathcal{S}$ & $\mathcal{S}$ \\
\hline
 $?$ & $?$ & $\mathcal{S}$ & $?$ & $?$ & $\mathcal{S}$ & $\mathcal{S}$ & $\mathcal{S}$ \\
\hline
 $?$ & $?$ & $\mathcal{S}$ & $?$ & $?$ & $\mathcal{S}$ & $\mathcal{S}$ & $\mathcal{S}$ \\
\hline
$?$ & $?$ & $\mathcal{S}$ & $?$  & $?$ & $\mathcal{S}$ & $\mathcal{S}$ & $\mathcal{S}$ \\
\hline
\end{tabular}
\end{center}
\caption{the 2th-order integral property of Rijndael-256} \label{256_2}
\end{figure}

We have found an other 4-round integral property of 2th-order as shown in figure \ref{256_2}. Using computer simulations, we have found 42 3th-order integral properties and 48 2th-order integral property (essentially the shifted ones).

\begin{figure}[ht]
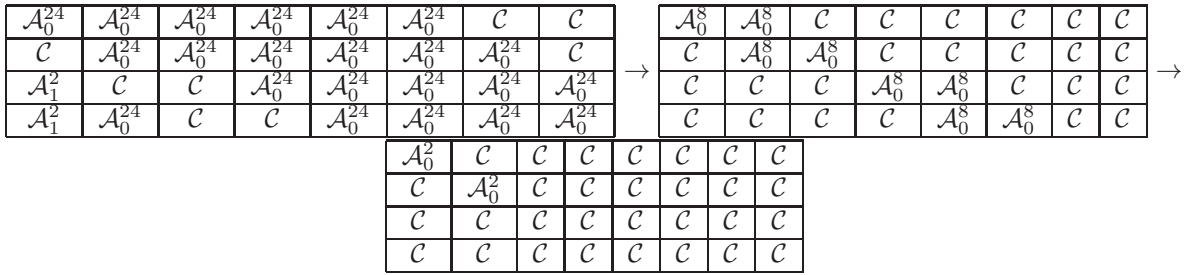

\begin{center}
\begin{tabular}{|c|c|c|c|c|c|c|c|}
\hline
$\mathcal{A}^{24}_0$ & $\mathcal{A}^{24}_0$ & $\mathcal{A}^{24}_0$ & $\mathcal{A}^{24}_0$ & $\mathcal{A}^{24}_0$ & $\mathcal{A}^{24}_0$ & $\mathcal{C}$ & $\mathcal{C}$ \\
\hline
 $\mathcal{C}$ & $\mathcal{A}^{24}_0$ & $\mathcal{A}^{24}_0$ & $\mathcal{A}^{24}_0$ & $\mathcal{A}^{24}_0$ & $\mathcal{A}^{24}_0$ & $\mathcal{A}^{24}_0$ & $\mathcal{C}$\\
\hline
$\mathcal{A}^2_1$ & $\mathcal{C}$ & $\mathcal{C}$ & $\mathcal{A}^{24}_0$ & $\mathcal{A}^{24}_0$ & $\mathcal{A}^{24}_0$ & $\mathcal{A}^{24}_0$ & $\mathcal{A}^{24}_0$ \\
\hline
$\mathcal{A}^2_1$ & $\mathcal{A}^{24}_0$ & $\mathcal{C}$  & $\mathcal{C}$ & $\mathcal{A}^{24}_0$ & $\mathcal{A}^{24}_0$ & $\mathcal{A}^{24}_0$ & $\mathcal{A}^{24}_0$ \\
\hline
\end{tabular}
$\rightarrow$
\begin{tabular}{|c|c|c|c|c|c|c|c|}
\hline
$\mathcal{A}^8_0$ & $\mathcal{A}^8_0$ & $\mathcal{C}$ & $\mathcal{C}$ & $\mathcal{C}$ & $\mathcal{C}$ & $\mathcal{C}$ & $\mathcal{C}$ \\
\hline
 $\mathcal{C}$ & $\mathcal{A}^8_0$ & $\mathcal{A}^8_0$ & $\mathcal{C}$ & $\mathcal{C}$ & $\mathcal{C}$ & $\mathcal{C}$ & $\mathcal{C}$\\
\hline
$\mathcal{C}$ & $\mathcal{C}$ & $\mathcal{C}$ & $\mathcal{A}^8_0$ & $\mathcal{A}^8_0$ & $\mathcal{C}$ & $\mathcal{C}$ & $\mathcal{C}$ \\
\hline
$\mathcal{C}$ & $\mathcal{C}$ & $\mathcal{C}$  & $\mathcal{C}$ & $\mathcal{A}^8_0$ & $\mathcal{A}^8_0$ & $\mathcal{C}$ & $\mathcal{C}$ \\
\hline
\end{tabular}
$\rightarrow$
\begin{tabular}{|c|c|c|c|c|c|c|c|}
\hline
$\mathcal{A}^2_0$ & $\mathcal{C}$ & $\mathcal{C}$ & $\mathcal{C}$ & $\mathcal{C}$ & $\mathcal{C}$ & $\mathcal{C}$ & $\mathcal{C}$ \\
\hline
 $\mathcal{C}$ & $\mathcal{A}^2_0$ & $\mathcal{C}$ & $\mathcal{C}$ & $\mathcal{C}$ & $\mathcal{C}$ & $\mathcal{C}$ & $\mathcal{C}$ \\
\hline
 $\mathcal{C}$ & $\mathcal{C}$ & $\mathcal{C}$ & $\mathcal{C}$ & $\mathcal{C}$ & $\mathcal{C}$ & $\mathcal{C}$ & $\mathcal{C}$ \\
\hline
$\mathcal{C}$ & $\mathcal{C}$ & $\mathcal{C}$ & $\mathcal{C}$  & $\mathcal{C}$ & $\mathcal{C}$ & $\mathcal{C}$ & $\mathcal{C}$ \\
\hline
\end{tabular}
\end{center}
\caption{Extension of Rijndael-256 by two rounds at the beginning using a 24th-order integral} \label{256_ext}
\end{figure}

As previously done, this 2th-order four-round property could be extended by one round at the beginning using a 8th-order integral (considering that it represents $2^{48}$ copies of the 2th-order four round integral) as previously described and by two rounds at the beginning using a 24th-order integral as done in \cite{DBLP:conf/mycrypt/NakaharaFP05} and as shown in Figure \ref{256_ext}.

Thus, we obtain first a four-round distinguisher that uses $2^{16}$ plaintexts testing if the sum taken over all initial values of a particular byte belonging to the third, to the sixth, to the seventh or to the eighth column is equal to zero. We also obtain a five-round distinguisher that uses $2^{64}$ plaintexts testing if the same sum taken over the $2^{64}$ values is also equal to zero. The six-round distinguisher that uses $2^{192}$ plaintexts is the same even if the corresponding memory complexity is here unreachable.


\subsubsection{Rijndael-224}
\begin{figure}[ht]
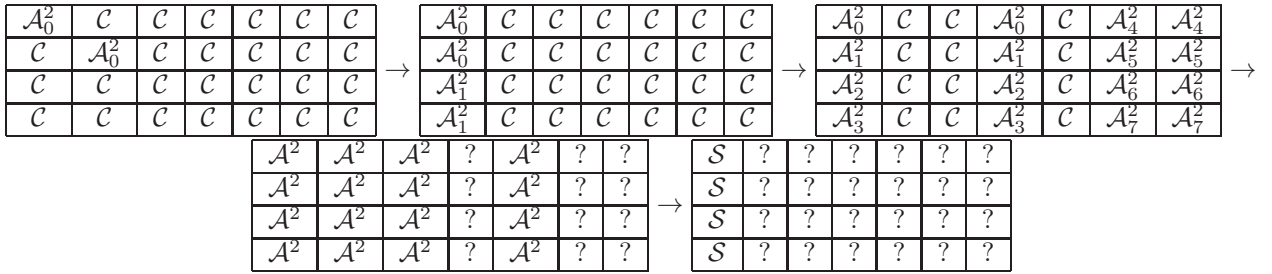

\begin{center}
\begin{tabular}{|c|c|c|c|c|c|c|}
\hline
$\mathcal{A}^2_0$ & $\mathcal{C}$ & $\mathcal{C}$ & $\mathcal{C}$ & $\mathcal{C}$ & $\mathcal{C}$ & $\mathcal{C}$ \\
\hline
 $\mathcal{C}$ & $\mathcal{A}^2_0$ & $\mathcal{C}$ & $\mathcal{C}$ & $\mathcal{C}$ & $\mathcal{C}$ & $\mathcal{C}$ \\
\hline
 $\mathcal{C}$ & $\mathcal{C}$ & $\mathcal{C}$ & $\mathcal{C}$ & $\mathcal{C}$ & $\mathcal{C}$ & $\mathcal{C}$ \\
\hline
$\mathcal{C}$ & $\mathcal{C}$ & $\mathcal{C}$ & $\mathcal{C}$  & $\mathcal{C}$ & $\mathcal{C}$ & $\mathcal{C}$ \\
\hline
\end{tabular}
$\rightarrow$
\begin{tabular}{|c|c|c|c|c|c|c|}
\hline
$\mathcal{A}^2_0$ & $\mathcal{C}$ & $\mathcal{C}$ & $\mathcal{C}$ & $\mathcal{C}$ & $\mathcal{C}$ & $\mathcal{C}$ \\
\hline
 $\mathcal{A}^2_0$ & $\mathcal{C}$ & $\mathcal{C}$ & $\mathcal{C}$ & $\mathcal{C}$ & $\mathcal{C}$ & $\mathcal{C}$ \\
\hline
$\mathcal{A}^2_1$ & $\mathcal{C}$ & $\mathcal{C}$ & $\mathcal{C}$ & $\mathcal{C}$ & $\mathcal{C}$ & $\mathcal{C}$ \\
\hline
$\mathcal{A}^2_1$ & $\mathcal{C}$ & $\mathcal{C}$  & $\mathcal{C}$ & $\mathcal{C}$ & $\mathcal{C}$ & $\mathcal{C}$ \\
\hline
\end{tabular}
$\rightarrow$
\begin{tabular}{|c|c|c|c|c|c|c|}
\hline
$\mathcal{A}^2_0$ & $\mathcal{C}$ & $\mathcal{C}$  & $\mathcal{A}^2_0$ & $\mathcal{C}$  & $\mathcal{A}^2_4$ & $\mathcal{A}^2_4$   \\
\hline
 $\mathcal{A}^2_1$ & $\mathcal{C}$ & $\mathcal{C}$&  $\mathcal{A}^2_1$ & $\mathcal{C}$ & $\mathcal{A}^2_5$ & $\mathcal{A}^2_5$  \\
\hline
$\mathcal{A}^2_2$ & $\mathcal{C}$ & $\mathcal{C}$ & $\mathcal{A}^2_2$ & $\mathcal{C}$ & $\mathcal{A}^2_6$ & $\mathcal{A}^2_6$   \\
\hline
$\mathcal{A}^2_3$ & $\mathcal{C}$ & $\mathcal{C}$  & $\mathcal{A}^2_3$ & $\mathcal{C}$ & $\mathcal{A}^2_7$ & $\mathcal{A}^2_7$   \\
\hline
\end{tabular}
$\rightarrow$
\begin{tabular}{|c|c|c|c|c|c|c|}
\hline
$\mathcal{A}^2$ & $\mathcal{A}^2$ & $\mathcal{A}^2$ & $?$ & $\mathcal{A}^2$ & $?$ & $?$  \\
\hline
 $\mathcal{A}^2$ & $\mathcal{A}^2$ & $\mathcal{A}^2$ & $?$ & $\mathcal{A}^2$ & $?$ & $?$  \\
\hline
 $\mathcal{A}^2$ & $\mathcal{A}^2$ & $\mathcal{A}^2$ & $?$ & $\mathcal{A}^2$ & $?$ & $?$  \\
\hline
$\mathcal{A}^2$ & $\mathcal{A}^2$ & $\mathcal{A}^2$ & $?$  & $\mathcal{A}^2$ & $?$ & $?$  \\
\hline
\end{tabular}
$\rightarrow$
\begin{tabular}{|c|c|c|c|c|c|c|}
\hline
 $\mathcal{S}$ & $?$ & $?$ & $?$ & $?$ & $?$ & $?$  \\
\hline
 $\mathcal{S}$ & $?$ & $?$ & $?$ & $?$ & $?$ & $?$  \\
\hline
$\mathcal{S}$ & $?$ & $?$ & $?$ & $?$ & $?$ & $?$ \\
\hline
$\mathcal{S}$ & $?$  & $?$ & $?$ & $?$ & $?$ & $?$ \\
\hline
\end{tabular}
\end{center}
\caption{Four-round 2th-order integral property of Rijndael-224} \label{224_1}
\end{figure}

In the same way, we have found a 2th-order 4-round integral property for Rijndael-224 as shown in figure \ref{224_1}. We have found 42 2th-order integral properties (essentially the shifted ones).

As previously done, this 2th-order four-round  property could be extended by one round at the beginning using a 8th-order integral (considering that it represents $2^{48}$ copies of the 2th-order four-round integral) as previously described and by two rounds at the beginning using a 24th-order integral.

Thus, we obtain first a four-round distinguisher that uses $2^{16}$ plaintexts testing if the sum taken over all initial values of a particular byte belonging to the first column is equal to zero. We also obtain a five-round distinguisher that uses $2^{64}$ plaintexts testing if the same sum taken over the $2^{64}$ values is also equal to zero. The six-round distinguisher that uses $2^{192}$ plaintexts is the same.


\subsubsection{Rijndael-192}
\begin{figure}[ht]
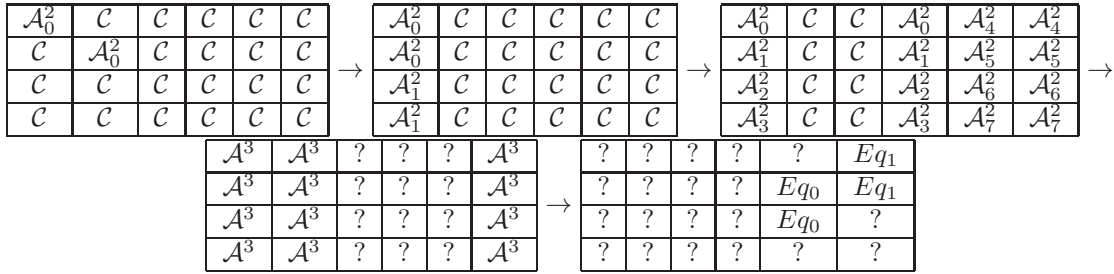

\begin{center}
\begin{tabular}{|c|c|c|c|c|c|}
\hline
$\mathcal{A}^2_0$ & $\mathcal{C}$ & $\mathcal{C}$ & $\mathcal{C}$ & $\mathcal{C}$ & $\mathcal{C}$  \\
\hline
 $\mathcal{C}$ & $\mathcal{A}^2_0$ & $\mathcal{C}$ & $\mathcal{C}$ & $\mathcal{C}$ & $\mathcal{C}$  \\
\hline
 $\mathcal{C}$ & $\mathcal{C}$ & $\mathcal{C}$ & $\mathcal{C}$ & $\mathcal{C}$ & $\mathcal{C}$  \\
\hline
$\mathcal{C}$ & $\mathcal{C}$ & $\mathcal{C}$ & $\mathcal{C}$  & $\mathcal{C}$ & $\mathcal{C}$  \\
\hline
\end{tabular}
$\rightarrow$
\begin{tabular}{|c|c|c|c|c|c|}
\hline
$\mathcal{A}^2_0$ & $\mathcal{C}$ & $\mathcal{C}$ & $\mathcal{C}$ & $\mathcal{C}$ & $\mathcal{C}$  \\
\hline
 $\mathcal{A}^2_0$ & $\mathcal{C}$ & $\mathcal{C}$ & $\mathcal{C}$ & $\mathcal{C}$ & $\mathcal{C}$  \\
\hline
$\mathcal{A}^2_1$ & $\mathcal{C}$ & $\mathcal{C}$ & $\mathcal{C}$ & $\mathcal{C}$ & $\mathcal{C}$  \\
\hline
$\mathcal{A}^2_1$ & $\mathcal{C}$ & $\mathcal{C}$  & $\mathcal{C}$ & $\mathcal{C}$ & $\mathcal{C}$  \\
\hline
\end{tabular}
$\rightarrow$
\begin{tabular}{|c|c|c|c|c|c|}
\hline
$\mathcal{A}^2_0$ & $\mathcal{C}$ & $\mathcal{C}$  & $\mathcal{A}^2_0$   & $\mathcal{A}^2_4$ & $\mathcal{A}^2_4$   \\
\hline
 $\mathcal{A}^2_1$ & $\mathcal{C}$ & $\mathcal{C}$&  $\mathcal{A}^2_1$  & $\mathcal{A}^2_5$ & $\mathcal{A}^2_5$  \\
\hline
$\mathcal{A}^2_2$ & $\mathcal{C}$ & $\mathcal{C}$ & $\mathcal{A}^2_2$  &  $\mathcal{A}^2_6$ & $\mathcal{A}^2_6$   \\
\hline
$\mathcal{A}^2_3$ & $\mathcal{C}$ & $\mathcal{C}$  & $\mathcal{A}^2_3$ &  $\mathcal{A}^2_7$ & $\mathcal{A}^2_7$   \\
\hline
\end{tabular}
$\rightarrow$
\begin{tabular}{|c|c|c|c|c|c|}
\hline
$\mathcal{A}^3$ & $\mathcal{A}^3$ & $?$ & $?$ & $?$ & $\mathcal{A}^3$   \\
\hline
 $\mathcal{A}^3$ & $\mathcal{A}^3$ & $?$ & $?$ & $?$ & $\mathcal{A}^3$  \\
\hline
 $\mathcal{A}^3$ & $\mathcal{A}^3$ & $?$ & $?$ & $?$ & $\mathcal{A}^3$  \\
\hline
$\mathcal{A}^3$ & $\mathcal{A}^3$ & $?$ & $?$  & $?$ & $\mathcal{A}^3$  \\
\hline
\end{tabular}
$\rightarrow$
\begin{tabular}{|c|c|c|c|c|c|}
\hline
 $?$ & $?$ & $?$ & $?$ & $?$ & $Eq_1$  \\
\hline
$?$ & $?$ & $?$ & $?$  & $Eq_0$ & $Eq_1$  \\
\hline
$?$ & $?$ & $?$ & $?$ & $Eq_0$ & $?$ \\
\hline
$?$  & $?$ & $?$ & $?$ & $?$ & $?$ \\
\hline
\end{tabular}
\end{center}
\caption{2th-order 4-round integral property of Rijndael-192} \label{192_1}
\end{figure}

In the same way, we have found a 2th-order 4-round integral property  for Rijndael-192 as shown in figure \ref{192_1}. This integral is different from the others because it implies that two particular sums are equals between them and not to zero. This particular property comes from the fact that the first term $Eq_0$ is a linear combination of 4 particular terms of the previous round. In those words, three are balanced (i.e. the complete sum at the end is equal to 0) and the last one comes from a `$?$' of the previous round. Thus, we obtain in fact the simple sum of the word `$?$', more precisely the sum becomes after the Mixcolumns $\bigoplus_{v \in \mathcal{A}^2_0} 01 \cdot `?' \oplus 0$ where $0$ is the null sum taken over the three balanced bytes. Then, notice that the second term $Eq_0$ is computed from exactly the same 4 words and that the coefficient of the Mixcolumns applied to the same word `$?$' is also $01$. Thus, we obtain two equal sums. In fact, on this particular column, we obtain 6 possible equalities up to the MixColumns coefficient.

\begin{figure}[ht]
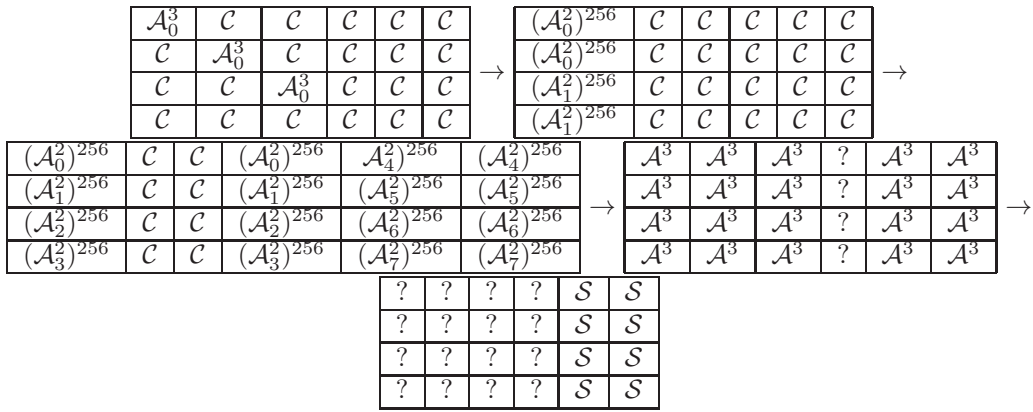

\begin{center}
\begin{tabular}{|c|c|c|c|c|c|}
\hline
$\mathcal{A}^3_0$ & $\mathcal{C}$ & $\mathcal{C}$ & $\mathcal{C}$ & $\mathcal{C}$ & $\mathcal{C}$  \\
\hline
 $\mathcal{C}$ & $\mathcal{A}^3_0$ & $\mathcal{C}$ & $\mathcal{C}$ & $\mathcal{C}$ & $\mathcal{C}$  \\
\hline
 $\mathcal{C}$ & $\mathcal{C}$ & $\mathcal{A}^3_0$ & $\mathcal{C}$ & $\mathcal{C}$ & $\mathcal{C}$  \\
\hline
$\mathcal{C}$ & $\mathcal{C}$ & $\mathcal{C}$ & $\mathcal{C}$  & $\mathcal{C}$ & $\mathcal{C}$  \\
\hline
\end{tabular}
$\rightarrow$
\begin{tabular}{|c|c|c|c|c|c|}
\hline
$(\mathcal{A}^2_0)^{256}$ & $\mathcal{C}$ & $\mathcal{C}$ & $\mathcal{C}$ & $\mathcal{C}$ & $\mathcal{C}$  \\
\hline
 $(\mathcal{A}^2_0)^{256}$ & $\mathcal{C}$ & $\mathcal{C}$ & $\mathcal{C}$ & $\mathcal{C}$ & $\mathcal{C}$  \\
\hline
$(\mathcal{A}^2_1)^{256}$ & $\mathcal{C}$ & $\mathcal{C}$ & $\mathcal{C}$ & $\mathcal{C}$ & $\mathcal{C}$  \\
\hline
$(\mathcal{A}^2_1)^{256}$ & $\mathcal{C}$ & $\mathcal{C}$  & $\mathcal{C}$ & $\mathcal{C}$ & $\mathcal{C}$  \\
\hline
\end{tabular}
$\rightarrow$
\begin{tabular}{|c|c|c|c|c|c|}
\hline
$(\mathcal{A}^2_0)^{256}$ & $\mathcal{C}$ & $\mathcal{C}$  & $(\mathcal{A}^2_0)^{256}$   & $\mathcal{A}^2_4)^{256}$ & $(\mathcal{A}^2_4)^{256}$   \\
\hline
 $(\mathcal{A}^2_1)^{256}$ & $\mathcal{C}$ & $\mathcal{C}$&  $(\mathcal{A}^2_1)^{256}$  & $(\mathcal{A}^2_5)^{256}$ & $(\mathcal{A}^2_5)^{256}$  \\
\hline
$(\mathcal{A}^2_2)^{256}$ & $\mathcal{C}$ & $\mathcal{C}$ & $(\mathcal{A}^2_2)^{256}$  &  $(\mathcal{A}^2_6)^{256}$ & $(\mathcal{A}^2_6)^{256}$   \\
\hline
$(\mathcal{A}^2_3)^{256}$ & $\mathcal{C}$ & $\mathcal{C}$  & $(\mathcal{A}^2_3)^{256}$ &  $(\mathcal{A}^2_7)^{256}$ & $(\mathcal{A}^2_7)^{256}$   \\
\hline
\end{tabular}
$\rightarrow$
\begin{tabular}{|c|c|c|c|c|c|}
\hline
$\mathcal{A}^3$ & $\mathcal{A}^3$ & $\mathcal{A}^3$ & $?$ & $\mathcal{A}^3$ & $\mathcal{A}^3$   \\
\hline
 $\mathcal{A}^3$ & $\mathcal{A}^3$ & $\mathcal{A}^3$ & $?$ & $\mathcal{A}^3$ & $\mathcal{A}^3$  \\
\hline
 $\mathcal{A}^3$ & $\mathcal{A}^3$ & $\mathcal{A}^3$ & $?$ & $\mathcal{A}^3$ & $\mathcal{A}^3$  \\
\hline
$\mathcal{A}^3$ & $\mathcal{A}^3$ & $\mathcal{A}^3$ & $?$  & $\mathcal{A}^3$ & $\mathcal{A}^3$  \\
\hline
\end{tabular}
$\rightarrow$
\begin{tabular}{|c|c|c|c|c|c|}
\hline
 $?$ & $?$ & $?$ & $?$ & $\mathcal{S}$ & $\mathcal{S}$  \\
\hline
$?$ & $?$ & $?$ & $?$  & $\mathcal{S}$ & $\mathcal{S}$  \\
\hline
$?$ & $?$ & $?$ & $?$ & $\mathcal{S}$ & $\mathcal{S}$ \\
\hline
$?$  & $?$ & $?$ & $?$ & $\mathcal{S}$ & $\mathcal{S}$ \\
\hline
\end{tabular}
\end{center}
\caption{3th-order 4-round integral property of Rijndael-192} \label{192_2}
\end{figure}

We also have found a 3th-order 4-round integral property for Rijndael-192 as shown in figure \ref{192_2}. We have found 42 2th-order integral properties (essentially the shifted ones).

As previously done, the 2th-order four-round integral could be extended by one round at the beginning using a 8th-order integral (considering that it represents $2^{48}$ copies of the 2th-order four-round integral) testing if the sums of two particular bytes are equals and by two rounds at the beginning using a 24th-order integral. In this case, we obtain exactly the same distinguisher than the ones described in the previous subsections.

In the case of the 3th-order four-round integral, it could be easily extended by one round at the beginning using a 12th-order integral (considering that it represents $2^{72}$ copies of the 3th-order four-round integral). But if we try to add one more round at the beginning, we need to consider the entire codebook of Rijndael-192, what is not possible (because, as mentioned in \cite{DBLP:conf/fse/FergusonKLSSWW00}, in this case in the key bytes search even the wrong keys will yield to zero when summing over all $2^{192}$ encryptions because Rijndael-192 is a permutation). Thus, in this case, we need to use the herd technique proposed in \cite{DBLP:conf/fse/FergusonKLSSWW00} and detailed in the Subsection \ref{herd}.

In conclusion, we obtain first a four-round distinguisher that uses $2^{24}$ plaintexts testing if the sum taken over all initial values of a particular byte belonging to the fifth or the sixth column is equal to zero. We also obtain a five-round distinguisher that uses $2^{92}$ plaintexts testing if the same sum taken over the $2^{92}$ values is also equal to zero. 


\subsubsection{Rijndael-160}

\begin{figure}[ht]
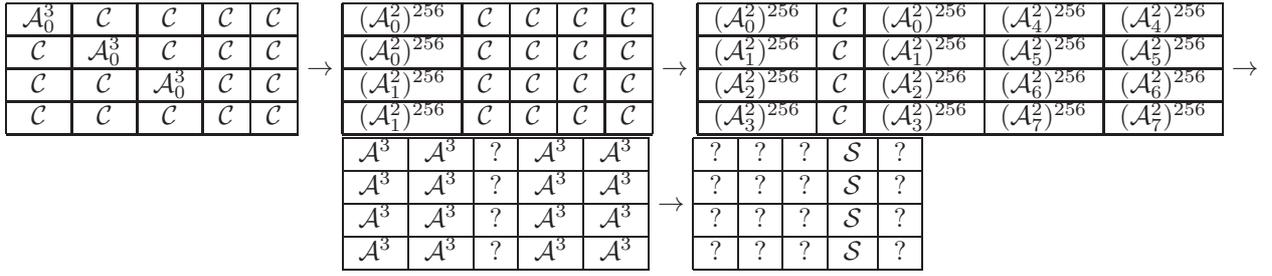

\begin{center}
\begin{tabular}{|c|c|c|c|c|}
\hline
$\mathcal{A}^3_0$ & $\mathcal{C}$ & $\mathcal{C}$ & $\mathcal{C}$ & $\mathcal{C}$   \\
\hline
 $\mathcal{C}$ & $\mathcal{A}^3_0$ & $\mathcal{C}$ & $\mathcal{C}$ & $\mathcal{C}$   \\
\hline
 $\mathcal{C}$ & $\mathcal{C}$ & $\mathcal{A}^3_0$ & $\mathcal{C}$ & $\mathcal{C}$  \\
\hline
$\mathcal{C}$ & $\mathcal{C}$ & $\mathcal{C}$ & $\mathcal{C}$  & $\mathcal{C}$ \\
\hline
\end{tabular}
$\rightarrow$
\begin{tabular}{|c|c|c|c|c|}
\hline
$(\mathcal{A}^2_0)^{256}$ & $\mathcal{C}$ & $\mathcal{C}$ & $\mathcal{C}$ & $\mathcal{C}$  \\
\hline
 $(\mathcal{A}^2_0)^{256}$ & $\mathcal{C}$ & $\mathcal{C}$ & $\mathcal{C}$ & $\mathcal{C}$ \\
\hline
$(\mathcal{A}^2_1)^{256}$ & $\mathcal{C}$ & $\mathcal{C}$ & $\mathcal{C}$ & $\mathcal{C}$  \\
\hline
$(\mathcal{A}^2_1)^{256}$ & $\mathcal{C}$ & $\mathcal{C}$  & $\mathcal{C}$ & $\mathcal{C}$ \\
\hline
\end{tabular}
$\rightarrow$
\begin{tabular}{|c|c|c|c|c|}
\hline
$(\mathcal{A}^2_0)^{256}$ & $\mathcal{C}$ & $(\mathcal{A}^2_0)^{256}$   & $(\mathcal{A}^2_4)^{256}$ & $(\mathcal{A}^2_4)^{256}$   \\
\hline
 $(\mathcal{A}^2_1)^{256}$  & $\mathcal{C}$&  $(\mathcal{A}^2_1)^{256}$  & $(\mathcal{A}^2_5)^{256}$ & $(\mathcal{A}^2_5)^{256}$  \\
\hline
$(\mathcal{A}^2_2)^{256}$  & $\mathcal{C}$ & $(\mathcal{A}^2_2)^{256}$  & $(\mathcal{A}^2_6)^{256}$ & $(\mathcal{A}^2_6)^{256}$   \\
\hline
$(\mathcal{A}^2_3)^{256}$  & $\mathcal{C}$  & $(\mathcal{A}^2_3)^{256}$ & $(\mathcal{A}^2_7)^{256}$ & $(\mathcal{A}^2_7)^{256}$   \\
\hline
\end{tabular}
$\rightarrow$
\begin{tabular}{|c|c|c|c|c|}
\hline
  $\mathcal{A}^3$ & $\mathcal{A}^3$ & $?$ & $\mathcal{A}^3$ & $\mathcal{A}^3$   \\
\hline
 $\mathcal{A}^3$ & $\mathcal{A}^3$ & $?$ & $\mathcal{A}^3$ & $\mathcal{A}^3$  \\
\hline
 $\mathcal{A}^3$ & $\mathcal{A}^3$ & $?$ & $\mathcal{A}^3$ & $\mathcal{A}^3$  \\
\hline
$\mathcal{A}^3$ & $\mathcal{A}^3$ & $?$  & $\mathcal{A}^3$ & $\mathcal{A}^3$  \\
\hline
\end{tabular}
$\rightarrow$
\begin{tabular}{|c|c|c|c|c|}
\hline
 $?$ & $?$ & $?$ & $\mathcal{S}$ & $?$  \\
\hline
 $?$ & $?$ & $?$  & $\mathcal{S}$ & $?$  \\
\hline
 $?$ & $?$ & $?$ & $\mathcal{S}$ & $?$ \\
\hline
 $?$ & $?$ & $?$ & $\mathcal{S}$ & $?$ \\
\hline
\end{tabular}
\end{center}
\caption{Integral property of Rijndael-160} \label{160_1}
\end{figure}

We also have found a 3th-order 4-round integral property for Rijndael-160 as shown in figure \ref{160_1}. We have found 42 3th-order integral properties (essentially the shifted ones).

Thus, we could easily extend by one round at the beginning this integral using a 12th-order integral (considering that it represents $2^{72}$ copies of the 3th-order four-round integral). We could not add one more round at the beginning, for the same reasons than the ones given in the case of Rijndael-192. Thus, in this case, we need to use the herd technique proposed in \cite{DBLP:conf/fse/FergusonKLSSWW00}.

Thus, we obtain first a four-round distinguisher that uses $2^{24}$ plaintexts testing if the sum taken over all initial values of a particular byte belonging to the fourth column is equal to zero. We also obtain a five-round distinguisher that uses $2^{92}$ plaintexts testing if the same sum taken over the $2^{92}$ values is also equal to zero. 

%
%

\section{The proposed attacks} \label{attacks}
We could exploit the 4, 5 and 6-round integral properties previously described to mount elementary attacks against 6, 7 and 8 rounds versions of Rijndael-$b$ using the partial sums technique described in \cite{DBLP:conf/fse/FergusonKLSSWW00} to add two rounds at the end. To attack the 8-round versions of Rijndael-192 and Rijndael-160, we introduce the herd technique also described in \cite{DBLP:conf/fse/FergusonKLSSWW00}.

\subsection{The partial sums technique}
We could extend the previous 5-round and 6-round distinguishers by adding two rounds at the end  using the partial sums technique introduced in \cite{DBLP:conf/fse/FergusonKLSSWW00}. We describe here the original attack and then directly apply it to our case.

This extension works in the original paper on a 6 rounds version of the AES and looks at a particular byte of $A^{(5)}$ to test the 4th-order integral property described in Figure \ref{int4} and how it relates to the ciphertext.  First, the authors rewrite the cipher slightly by putting the AddRoundKey before the MixColumns in round 5. Instead of applying MixColumns and then adding $K_5$, they first add in $K'_5$, which is a linear combination of four bytes of $K_5$, and then apply MixColumns. Under this assumption, it is easy to see that any byte of $A^{(5)}$ depends on the ciphertext, on four bytes of $K_6$ and one byte of $K'_5$ considering that the sixth round is the last one and does not contain a MixColumns operation. Then, only the five key bytes of the two last rounds remain unknowns. 

Moreover, the authors improve the complexity of their attack using a technique called ``partial sums'' to sequentially decipher the two last rounds according to the values of the five unknown key bytes. They first compute from the $i$-th ciphertext $c_i$ the following partial sums: $\forall k \in \{0, \cdots, 3\}, \; \; x_k := \sum_{j=0}^k{S_j \left[ c_{i,j} \oplus k_j \right]}$where $S_0$, $S_1$, $S_2$, $S_3$  represent the inverse of the S-box $S$ multiplied by a component of InvMixColumns, $c_{i,j}$ the byte number $j$ of $c_i$; $k_0, \cdots, k_3$ the four bytes of $K_6$. Note that the searched value at the end of the 4 rounds is thus $a^{(5)}_{i,j} = S^{-1}[x_3 \oplus k_4]$ where $k_4$ is the implied byte of $K'_5$.

They use the transformation $(c_0,c_1,c_2,c_3) \rightarrow (x_k,c_{k+1}$ $, \cdots, c_3)$ to sequentially determine the different values of $k_k$ and to share the global computation into 4 steps of key bytes search with $2^{48}$ operations for each one corresponding with $2^{50}$ S-box lookups for each set of $2^{32}$ ciphertexts (see \cite{DBLP:conf/fse/FergusonKLSSWW00} for the details of the complexities). To discard false alarms (i.e. bad keys that pass the test), they need to repeat this process on 6 different sets with $2^{32}$ elements. Then, the general complexity of the partial sums attacks against a 6 rounds version of the AES is about $2^{44}$ encryptions (considering that $2^{8}$ S-box applications are roughly equivalent with one trial encryption) using $6 \cdot 2^{32}$ plaintexts.

We could directly apply this technique to all the versions of Rijndael-$b$ to recover 5 particular key bytes of the two last rounds using 6 different sets of plaintexts. We sum up the corresponding results in Table \ref{ps}. Note also that when looking at the 2th-order four-round integral of Rijndael-192, one needs to guess in parallel $2 \times 5$ key bytes. The partial sums technique could however be applied 2 times but all the first 5 guessed key bytes must be stored. We thus increase the required memory. Note also, as done in \cite{DBLP:conf/fse/FergusonKLSSWW00}, that we could add at the end of the two rounds added using the partial sums technique a last round guessing 4 particular columns of the last subkey. In this case, we perform an exhaustive search on 16 subkey bytes whereas the 5 other key bytes are determined using always the partial sums technique. The corresponding results are also given in Table \ref{ps}.

\begin{table}[ht]
\begin{center}
\begin{tabular}{ccccccc}
\hline
Cipher &  nb  & Key & Data & Time & Memory & Attack\\
& rounds & sizes & & Complexity & & \\
\hline
Rijndael-256 & 6 & (all) & $2 \cdot 2^{16}$ CP & $2^{32}$ &  $2^{16}$ & 2th-order integral \\
& 7 & (all) & $6 \cdot 2^{64}$ CP & $2^{80}$ & $2^{64}$ & 8th-order integral \\
 & 8 & $> 192$ & $6 \cdot 2^{192}$ CP & $2^{208}$ & $2^{192}$ & 24th-order integral \\
 & 8 &  (192) & $19 \cdot 2^{64}$ CP & $2^{191}$ & $2^{64}$ & 8th-order integral \\
 & 8 & (256) & $21 \cdot 2^{64}$ CP & $2^{207}$ & $2^{64}$ & 8th-order integral \\
 \hline
 Rijndael-224 & 6 & (all) & $2 \cdot 2^{16}$ CP & $2^{32}$ &  $2^{16}$ & 2th-order integral \\
& 7 & (all) & $6 \cdot 2^{64}$ CP & $2^{80}$ & $2^{64}$ & 8th-order integral \\
 & 8 & $> 192$ & $6 \cdot 2^{192}$ CP & $2^{208}$ & $2^{192}$ & 24th-order integral \\
  & 8 &  (192) & $19 \cdot 2^{64}$ CP & $2^{191}$ & $2^{64}$ & 8th-order integral \\
 & 8 & (256) & $21 \cdot 2^{64}$ CP & $2^{207}$ & $2^{64}$ & 8th-order integral \\
 \hline
 Rijndael-192 & 6 & (all) & $2 \cdot 2^{16}$ CP & $2^{33}$ &  $2^{16}$  & 2th-order integral\\
& 7 & (all) & $6 \cdot 2^{64}$ CP & $2^{81}$ & $2^{104}$ & 8th-order integral \\
& 6 & (all) & $2 \cdot 2^{24}$ CP & $2^{40}$ &  $2^{24}$  & 3th-order integral\\
& 7 & (all) & $6 \cdot 2^{92}$ CP & $2^{108}$ & $2^{92}$ & 12th-order integral \\
& 8 & (256) & $21 \cdot 2^{64}$ CP & $2^{208}$ & $2^{104}$ & 8th-order integral \\
\hline
Rijndael-160  & 6 & (all) & $2 \cdot 2^{24}$ CP & $2^{40}$ &  $2^{24}$ & 3th-order integral \\
& 7 & (all) & $6 \cdot 2^{92}$ CP & $2^{108}$ & $2^{92}$  & 12th-order integral \\
\hline
\end{tabular}

\caption{Summary of Attacks on Rijndael-$b$ using the partial sums technique} \label{ps}
\end{center}
\end{table}
\vspace{-4mm}

\subsection{The herd technique for Rijndael-192 and Rijndael-160} \label{herd}
In \cite{DBLP:conf/fse/FergusonKLSSWW00}, the authors develop a technique to improve their 6 rounds AES attack by adding one round at the beginning. This new attack require naively the entire codebook of $2^{128}$ known plaintexts that could be divided into $2^{96}$ packs of $2^{32}$ plaintexts/ciphertexts that represent $2^{24}$ first-order integrals with one active byte after two rounds. But this property could not be directly exploited because in this case even the wrong keys pass the test due to the bijective behavior of the cipher.

Instead, they use a particular byte at the end of the first round, say $a^{(2)}_{a,b}$ different from the four bytes implied in the integral with a fixed value $x$. With $a^{(2)}_{a,b}=x$, they obtain a set of $2^{120}$ possible encryptions composed of $2^{88}$ packs, where each pack contains $2^{24}$ 4th-order integrals. They call this structure with $2^{120}$ elements a \textit{herd}. If they sum up values on a herd, then the integral property is only preserved for the correct key. 

Thus, they notice that this particular byte $a^{(2)}_{a,b}$ depends on only four bytes of plaintext, say $(p_4, \cdots, p_7)$ and on four bytes of the key $K_0$. As done for the partial sums technique, they could share the key exhaustive search on the four key bytes of $K_0$ required to entirely determine the value of $a^{(1)}_{a,b}$ in a three-phase attack using $2^{64}$ counters $m_y$ for the first phase, $2^{32}$ counters $n_z$ for the second whereas the third phase filters information for key guesses. The attack works as follows: in the first phase, the counter $m_y$ is incremented at bit level according to the 64-bit value $y=(c_0, \cdots, c_3, p_4, \cdots, p_7)$; in the second phase, the four bytes of $K_0$ are guessed to compute $a^{(2)}_{a,b}$ and to share the counters into herds; then select a single herd and update $n_z$ by adding $z=(c_0, \cdots, c_3)$ for each $y$ that is in the good herd; in the third phase, guess the five key bytes of $K_7$ and of $K'_6$ to decrypt each $z$ to a single byte of $A^{(6)}$, sum this byte over all the $2^{32}$ values of $z$ (with multiplicities) and check for zero. This last phase must be repeated for each initial guess of the four bytes of $K_0$.

The first phase requires about $2^{120}$ trial encryptions and the rest of the attack has a negligible complexity compared to it (see \cite{DBLP:conf/fse/FergusonKLSSWW00} for some details about the attack complexity). Then, the total complexity of this attack is $2^{120}$ trial encryptions and $2^{64}$ bits of memory using $2^{128}$ chosen plaintexts. The authors provide another improvement of their attack remarking that the four plaintext bytes $(p_4, \cdots, p_7)$ and the four guessed key bytes of $K_0$ define four bytes of $A^{(2)}$. So they can create $2^{24}$ smaller herds with $2^{104}$ elements by fixing three more bytes of $A^{(2)}$ to reduce the plaintext requirements to $2^{128}-2^{119}$ texts. 

So, we could directly apply this attack against 8 rounds version of Rijndael-192 and Rijndael-160. The results are summed up in Table \ref{rr}. 

\vspace{-3mm}
\begin{table}[ht]
\begin{center}
\begin{tabular}{ccccccc}
\hline
Cipher &  nb  & Key & Data & Time & Memory & Attack\\
& rounds & sizes & & Complexity & & \\
\hline
Rijndael-192 & 8 & (256) & $2^{192}-2^{119}$ CP & $2^{192}-2^{119}$ & $2^{192}-2^{119}$ \\
Rijndael-160 & 8 & (256) & $2^{160}-2^{119}$ CP & $2^{160}-2^{119}$ & $2^{160}-2^{119}$ \\
\hline
\end{tabular}

\caption{Summary of Attacks on Rijndael-192 and on Rijndael-160 using the herd technique} \label{rr}
\end{center}
\end{table}
\vspace{-6mm}

\section{Conclusion} \label{conclusion}
In this paper, we have investigated new 4-round integral properties of Rijndael-$b$ for several $b$ values and then have built several deduced attacks up to 8 rounds. Note that those attacks are better than the ones described in \cite{DBLP:conf/isw/NakaharaP07}, in \cite{DBLP:conf/mycrypt/NakaharaFP05} and in \cite{DBLP:conf/fse/FergusonKLSSWW00} when applied to Rijndael-$b$ but do not improve the one proposed in \cite{DBLP:conf/africacrypt/GaliceM08} against Rijndael-256. 

However, we think that those new properties clearly improve the better results known about Rijndael-$b$ and highlight the Rijndael-$b$ behavior for integral cryptanalysis, noticing that the greater the number of columns is the lower the order of the integral is.   


\tableofcontents

\end{document}